\documentclass[journal]{IEEEtran}
\usepackage{graphicx}
\usepackage{algorithm}
\usepackage{algpseudocode}
\usepackage{xcolor}
\usepackage{multirow}
\usepackage{tabularx}
\usepackage{adjustbox}

\usepackage{amsmath}
\usepackage{amssymb}

\title{Online Alignment and Addition in Multi-Term Floating-Point Adders}
\author{Kosmas Alexandridis and Giorgos Dimitrakopoulos
\thanks{This work was supported by a Siemens EDA research grant to Democritus University of Thrace on ``High-level synthesis research for Systems-on-Chip''.}
\thanks{%
Kosmas Alexandridis and Giorgos Dimitrakopoulos are with the Department
of Electrical and Computer Engineering, Democritus University of Thrace, Xanthi, Greece.
E-mail: \{koalexan, dimitrak\}@ee.duth.gr}}

\begin{document}

\maketitle

\begin{abstract}
Multi-term floating-point addition appears in vector dot-product computations, matrix multiplications, and other forms of floating-point data aggregation. A critical step in multi-term floating point addition is the alignment of fractions of the floating-point terms
before adding them. Alignment is executed serially by identifying first the maximum of all exponents and then shifting 
the fraction of each term according to the difference of its exponent from the maximum one.
Contrary to common practice, this work proposes a new \emph{online} algorithm that splits the identification of the maximum exponent, the alignment shift for each fraction, and their addition to multiple fused incremental steps that can be computed in parallel. Each fused step is implemented by a new associative operator that allows the incremental alignment and addition for arbitrary number of operands. Experimental results show that employing the proposed align-and-add operators for the implementation of multi-term floating point adders can improve delay or 
save significant area and power. 
The achieved area and power savings range between 3\%--23\% and 4\%--26\%, respectively.
\end{abstract}


\section{Introduction}
Machine learning (ML) algorithms 
have been widespread in various application domains. 
Their efficient and accurate computation relies mostly on matrix multiplication kernels and floating-point (FP) arithmetic for data representation~\cite{miniFloat, ten-lessons}. 

The FP representations used in ML algorithms cover
IEEE-754 compliant formats as well as reduced-precision formats that use 16 or fewer bits in total, in an effort to balance numerical performance, and hardware and storage costs~\cite{bfloat, NIA-fp8}. 
In most cases, a FP number consists of three fields: the sign bit ($s$), the exponent ($e$), and the
fraction ($m$) and its value is given by 
$(-1)^s\times 1.m\times 2^{e-\text{bias}}$, where bias is a
constant that depends on the bit width of the exponent. 
Corner cases, such as not-a-number, infinity, or de-normals can be also encoded or skipped depending on the chosen format~\cite{NIA-fp8}.

To reduce the overhead of FP arithmetic when implementing vector-wide operations, designers have turned to fusing individual FP operations to more complex ones that implement the needed computation at once~\cite{yao-corr-round, swartz-3term-adder, swartz-DP4, reduced-FP-sys}. This approach allows alignment, normalization, and rounding steps to be shared among independent operations, ultimately resulting in more efficient hardware architectures. 

Multi-term addition, the core of fused operators, involves adding multiple FP numbers with potentially different exponents. To align the addends for addition, 
the fraction of each number is shifted according to the difference of its own exponent to the maximum exponent of all terms.
This serial dependency across fraction alignment and addition impacts negatively the overall hardware efficiency.

In this work, inspired by online softmax computation~\cite{online-softmax},
we propose a new \emph{online} approach for alignment and addition in multi-term FP adders. In this way, \emph{all serial dependencies} that traditionally characterize alignment and addition steps \emph{are removed} and maximum exponent calculation, as well as alignment and addition of fractions, are computed \emph{incrementally and in parallel}. 
In practice, alignment and addition is performed using trees built from the newly proposed align-and-add operators.

The experimental evaluation
shows that the proposed approach simplifies fundamentally the complexity of alignment and addition in multi-term FP adders. The corresponding hardware units that adopt the online alignment and addition paradigm, 
require significantly less area and power than traditional approaches. The area and power savings range between 3\%--23\% and 4\%--26\%, respectively, for various examined configurations. Also, when opting for high-speed implementations, they can also improve delay under the same number of pipeline stages. 

\section{Alignment and addition in multi-term floating point adders}
\label{s:baseline}

A high-level description of multi-term fused addition is shown in Algorithm~\ref{alg:add}. The input is an array of FP numbers $f_i$ and the output is their sum $S$. The algorithm begins by finding the exponent with the maximum value in step 1. Then, the fractions are aligned based on the difference of the local exponent and the maximum one (step 2). With the fractions aligned, the summation operation is performed in step 3.
The sum is normalized and rounded in step 4.

\begin{algorithm}[thb]
\caption{Multi-term fused floating point addition}\label{alg:add}
\hspace*{\algorithmicindent} \textbf{Input:} Floats $f_1, f_2, \ldots, f_N$ \\
\hspace*{\algorithmicindent} \textbf{Output: } $S = \sum_{i=1}^{N}{f_i}$
\begin{algorithmic}[1]
\State Find maximum exponent $e_{\max}=\max(e_1, e_2, \ldots, e_N)$
\State Align every fraction $1.m_i$ by shifting right by $e_{\max}-e_i$ positions    
\State Sum the aligned fractions $S=\sum_{i=1}^{N} \text{aligned}(1.m_i)$
\State Normalize and round $S$ to produce the final FP sum
\end{algorithmic}
\end{algorithm}    

The baseline implementation of alignment and addition (steps 1--3), which are the focus of this work, is detailed in Algorithm~\ref{alg:serial}.
The first loop corresponds to the first step in Algorithm~\ref{alg:add} that computes the maximum exponent and stores it in $\lambda_N$ at the end of the loop.
The second loop performs steps 2 and 3 of Algorithm~\ref{alg:add}; each fraction $m_i$ is aligned in line 5 and accumulated to a partial sum $o_i$ in line 6. To simplify presentation in Algorithm~\ref{alg:serial}, each fraction $1.m_i$ is denoted as $m_i$, which is assumed to be in signed (2's complement) form according to the sign $s_i$ of $f_i$.

\begin{algorithm}[h!]
\caption{Serial fraction alignment and addition}\label{alg:serial}
\begin{algorithmic}[1]
\For{$i\gets 1:N$}
\State $\lambda_i \gets \max(\lambda_{i-1}, e_i)$
\EndFor\Comment{Maximum exponent in $\lambda_N$}
\For{$i\gets 1:N$}
\State $am_i \gets m_i \gg (\lambda_N - e_i)$
\Comment{Alignment shift}
\State $o_i \gets o_{i-1} + am_i$ 
\Comment{Accumulate the aligned fraction}
\EndFor 
\State $S = o_N$
\end{algorithmic}
\end{algorithm}

\begin{figure}[t]
\centering
\includegraphics[width=0.75\columnwidth]{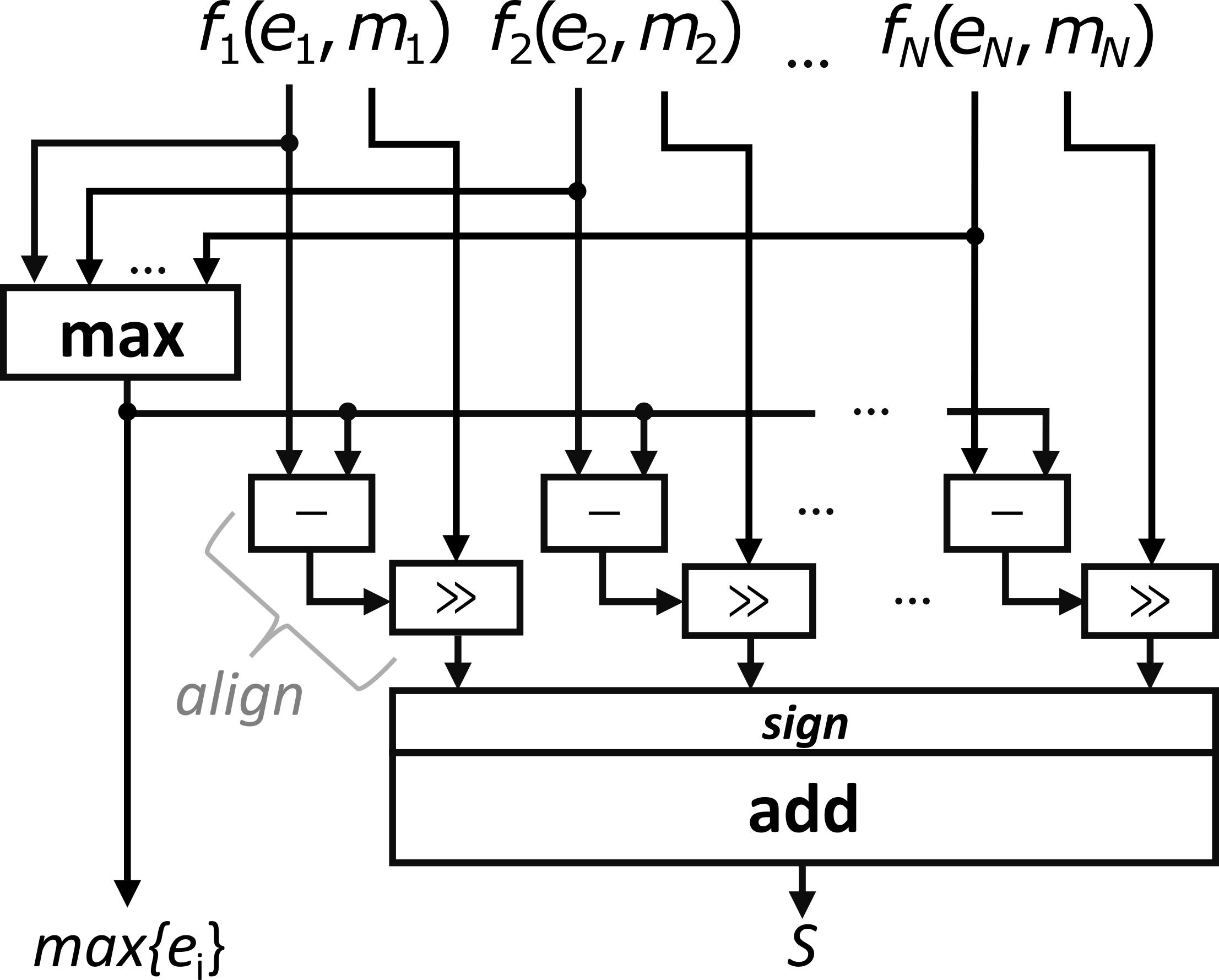}
\caption{Baseline approach for multi-term fraction alignment and addition.}
\label{f:serial-align-add}
\end{figure}

The two loops of Algorithm~\ref{alg:serial} cannot be merged. Thus, in hardware, each part is unrolled separately and the second loop can begin execution only after the first loop has computed the maximum exponent $\lambda_N$. This approach for alignment and addition is followed 
in the majority of hardware architectures for multi-term adders~\cite{intel-nervana, filippas, partial_align} and is shown in Fig.~\ref{f:serial-align-add}.

To reduce delay, other architectures perform fraction alignment based on the relative difference of exponents, and avoid the dependency to the maximum exponent that is computed in parallel~\cite{yao-corr-round, swartz-3term-adder, swartz-DP4, tenca}. However, in all cases, this concept is applied only for 3- or 4-term adders and cannot be generalized to arbitrary number of terms. This limitation is removed by the formulation in this work.
Other solutions, such as Kaul~\emph{et al.}~\cite{lg-align}, split the alignment of fractions into global and local alignment. 
In this way, computing exponent differences and the alignment shift are partially overlapped in time at the circuit level. However, still addition is performed separately in a following step.

Other approaches avoid the need for fraction alignment by mapping floating point accumulation to fixed-point arithmetic~\cite{dse-large-acc,large-acc-accelerator}. Effectively, alignment is performed implicitly when transforming FP numbers to their equivalent fixed-point integers. Such approaches are practical when accumulation is done in time. In this work, we focus on wide parallel architectures that perform addition in space. 

\section{Online alignment and addition}
\label{s:online-alg}
This work aims to fuse the serial alignment and addition steps into one combined step that would perform  maximum exponent calculation, alignment shift and addition incrementally and in parallel for various groups of inputs. Effectively, this transformation would allow us to merge the two separate loops of Algorithm~\ref{alg:serial} into one single loop.

To present the proposed algorithm for online alignment and addition, we first merge the shift and add operations, shown in lines 5 and 6 of Algorithm~\ref{alg:serial}, into one equation as follows:
\begin{equation}
o_i = o_{i-1} + m_i \gg (\lambda_N - e_i),\quad \text{with}\quad 
\lambda_N = \max_i\{e_i\}
\label{e:first}
\end{equation}
The right shift in~\eqref{e:first} can be equivalently written as a multiplication with a negative power of two, i.e., 
\begin{equation}
o_i = o_{i-1} + m_i\, 2^{-(\lambda_N - e_i)}.
\label{e:base-recursion}
\end{equation}
Fully unrolling~\eqref{e:base-recursion} we can write the final sum $o_N$ as follows:
\begin{equation}
o_N = o_{N-1}+o_{N-2}+\ldots +o_1=\sum_{i=1}^{N} m_i\, 2^{-(\lambda_N - e_i)}
\label{e:unroll}
\end{equation}

\subsection{Basic online algorithm for alignment and addition}
To remove the dependency to $\lambda_N$ for the computation of the final sum $o_N$ we define a new sequence $o'_{i}$ 
\begin{equation}
o'_{i} = \sum_{j=1}^{i} m_j 2^{-(\lambda_i - e_j)}\quad
\text{with }\, \lambda_i = \max(\lambda_{i-1}, e_i)
\label{e:alt-seq}
\end{equation}
Sequence~\eqref{e:alt-seq} has the interesting property that its last term $o'_N$ is equal to $o_N$ 
defined in~\eqref{e:unroll}. 
Beginning from~\eqref{e:alt-seq}, 
our goal is to form a recursive relation that would connect $o'_i$ to $o'_{i-1}$. 
Initially, in~\eqref{e:alt-seq}, we separate the $i$th term $m_i\, 2^{-(\lambda_i - e_i)}$ from the rest:
\begin{equation}
o'_i = \left ( \sum_{j=1}^{i-1} m_j\, 2^{-(\lambda_i - e_j)} \right ) + m_i\, 2^{-(\lambda_i - e_i)}\nonumber
\end{equation}
Then, inside the parenthesis, we add and subtract the helper term $\lambda_{i-1}$
\begin{equation}
o'_i =  \left ( \sum_{j=1}^{i-1} m_j\, 2^{-(\lambda_i - \lambda_{i-1} + \lambda_{i-1} - e_j)} \right ) + m_i\, 2^{-(\lambda_i - e_i)}\nonumber    
\end{equation}
Finally, we factor out the term $2^{-(\lambda_i - \lambda_{i-1})}$
\begin{equation}     
o'_i = \left ( \sum_{j=1}^{i-1}m_j\, 2^{-(\lambda_{i-1} - e_j)}\right ) 2^{-(\lambda_i - \lambda_{i-1})} + m_i\, 2^{-(\lambda_i - e_i)}\label{e:last-term}
\end{equation}
According to~\eqref{e:alt-seq}, the term left in the parenthesis corresponds to $o'_{i-1}$.
Thus, introducing $o'_{i-1}$
into~\eqref{e:last-term} we get the sought recursive relation:
\begin{align}
o'_i  & = o'_{i-1}\,2^{-(\lambda_i - \lambda_{i-1})} + m_i\, 2^{-(\lambda_i - e_i)}\label{e:main} \\
& \text{with }\, \lambda_i = \max(\lambda_{i-1}, e_i)\nonumber
\end{align}

\begin{algorithm}[thb]
\caption{Online fused fraction alignment and addition}\label{alg:online}
\begin{algorithmic}[1]
\For{$i\gets 1:N$}
\State $\lambda_i \gets \max(\lambda_{i-1}, e_i)$
\State $o'_i \gets o'_{i-1} \gg (\lambda_i - \lambda_{i-1}) + m_i \gg (\lambda_i - e_i)$
\EndFor
\State $S = o'_N$ 
\end{algorithmic}
\end{algorithm}

Remapping multiplications with negative powers of two back to equivalent right arithmetic shift operations, the recursive relation in~\eqref{e:main} can be equivalently expressed as follows:
\begin{equation}
o'_i = o'_{i-1} \gg (\lambda_i - \lambda_{i-1}) + m_i \gg (\lambda_i - e_i)
\label{e:main_w_shifts}
\end{equation}
This mapping to shift operations is valid since 
$\lambda_i$ is the maximum of $\lambda_{i-1}$ and $e_i$ and thus
the shift amounts $\lambda_i-\lambda_{i-1}$ and $\lambda_i-e_i$ in~\eqref{e:main_w_shifts}, are always greater or equal to zero.

Algorithm~\ref{alg:online} uses recursive relation~\eqref{e:main_w_shifts} to compute alignment and addition \emph{online}.
At each iteration, a local maximum exponent is identified that drives local alignment shifts and accumulation of the output sum.
Even if this fused align and add operation needs an extra subtraction and shift per iteration relative to Algorithm~\ref{alg:serial}, the experimental results show that it leads to more efficient unrolled and pipelined hardware implementations.

\subsection{Parallel computation of fraction alignment and addition}
The computation of the sum of aligned fractions $S$ and the identification of the maximum exponent can be performed in parallel using a new operator $\circledcirc$ that is defined as:
\begin{equation}
\begin{gathered}
\begin{bmatrix}
\lambda_i \\
o_i
\end{bmatrix}
\!\circledcirc\!
\begin{bmatrix}
\lambda_j \\
o_j
\end{bmatrix}
= \\
\begin{bmatrix}
\max(\lambda_i, \lambda_j) \\
o_i\! \gg\! \left ( \max(\lambda_i, \lambda_j) - \lambda_i\right ) + 
o_j\! \gg \!\left ( \max(\lambda_i, \lambda_j) - \lambda_j\right )
\end{bmatrix}
\label{e:operator}
\end{gathered}
\end{equation}
It can be shown by induction using a derivation similar to~\eqref{e:last-term} that
the final sum $S$ and the maximum exponent of a set of FP numbers can be computed using the newly defined operator $\circledcirc$ as follows:
\begin{equation}
\begin{bmatrix}
\max\{e_i\} \\
S
\end{bmatrix}=
\begin{bmatrix}
e_1 \\
m_1
\end{bmatrix}\circledcirc
\begin{bmatrix}
e_2 \\
m_2
\end{bmatrix}\circledcirc
\ldots
\circledcirc
\begin{bmatrix}
e_N \\
m_N
\end{bmatrix}
\end{equation}

Also, it can be proven that the operator $\circledcirc$ is associative since
\begin{equation}
\left (
\begin{bmatrix}
e_1\\ m_1
\end{bmatrix}
\!\circledcirc\!
\begin{bmatrix}
e_2\\ m_2
\end{bmatrix}
\right )
\!\circledcirc\!
\begin{bmatrix}
e_3\\ m_3
\end{bmatrix}\!=\!
\begin{bmatrix}
e_1\\ m_1
\end{bmatrix}
\!\circledcirc\!
\left (
\begin{bmatrix}
e_2\\ m_2
\end{bmatrix}
\!\circledcirc\!
\begin{bmatrix}
e_3\\ m_3
\end{bmatrix}
\right )
\end{equation}

\subsection{Hardware Organization of Alignment and Addition} 
\begin{figure}
\centering
\includegraphics[width=0.9\columnwidth]{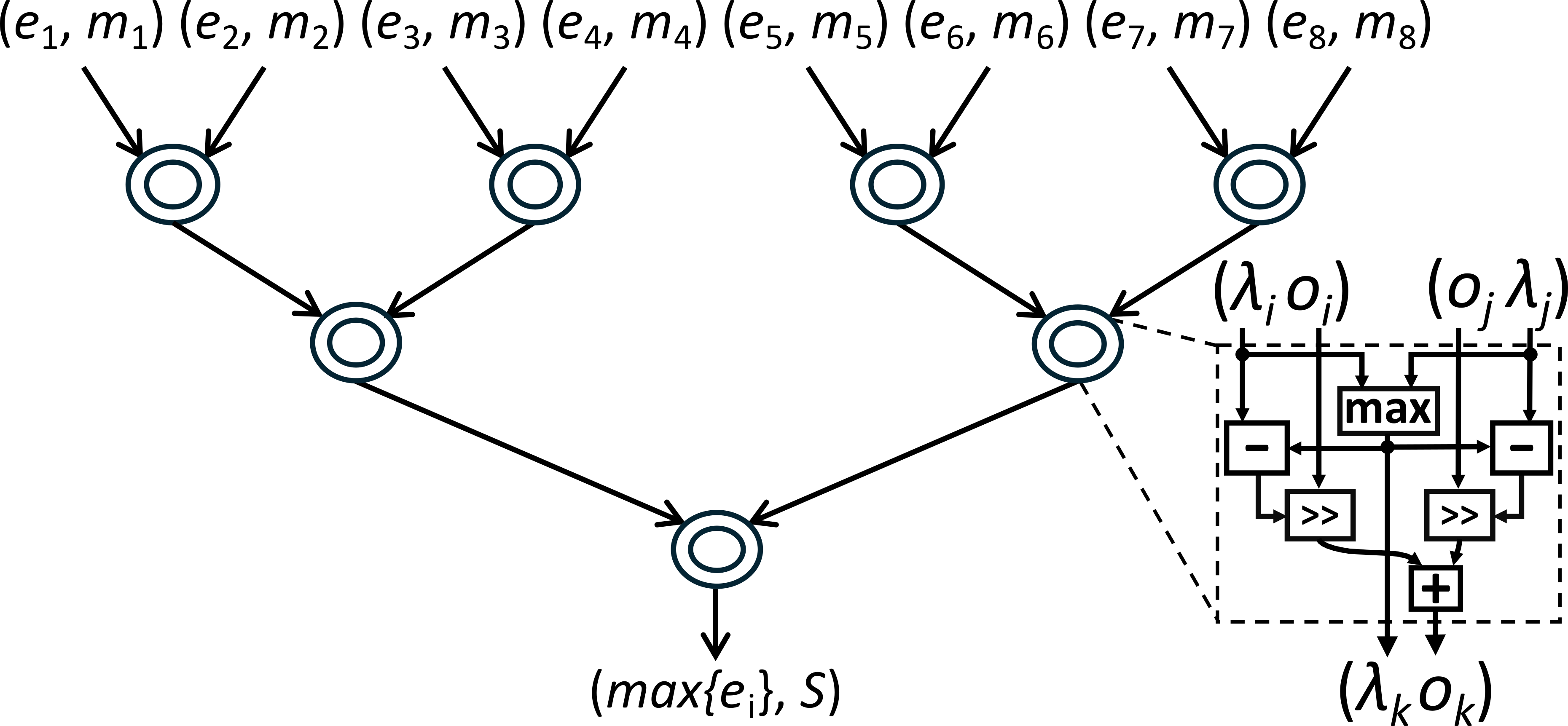} \\
{\small (a)} \\ \vspace{3mm}
\includegraphics[width=1.0\columnwidth]{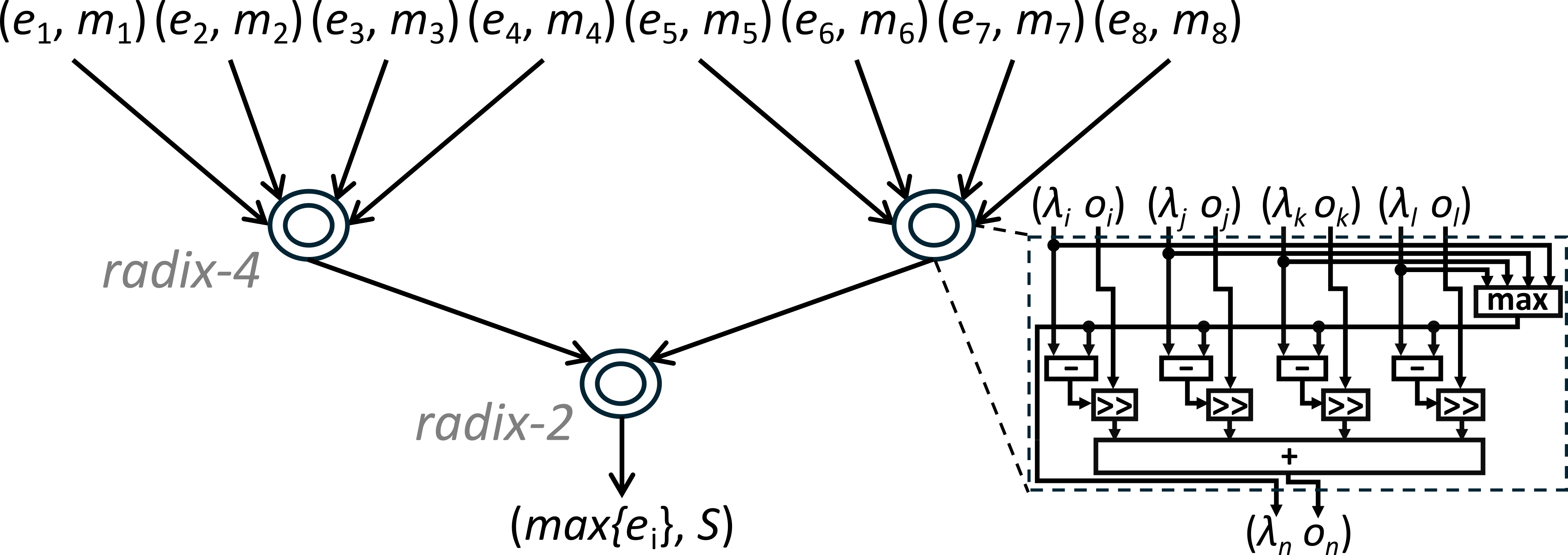} \\
{\small(b)} 
\caption{Tree-based organization of parallel alignment and addition for an 8-term floating point addition using (a) the radix-2 $\circledcirc$ operator in all nodes of the tree and (b) a mixture of radix-4 and radix-2 operators.}
\label{f:tree-align-add}
\end{figure}

Using the new associative operator $\circledcirc$, fraction alignment and addition can be performed using various hardware configurations. 
For instance, Fig.~\ref{f:tree-align-add}(a) depicts a binary-tree architecture of $\circledcirc$ operators. Following the definition of the $\circledcirc$ operator in~\eqref{e:operator}, at each node of the tree, the local maximum exponent is identified first and in turn
drives local fraction alignment and addition.

The $\circledcirc$ operator can be generalized to higher radices as well. Fig.~\ref{f:tree-align-add}(b) shows an example of an 8-term alignment and addition using a mixture of radix-4 and radix-2 operators. Radix-4 operators are used in the first level and a radix-2 operator at the last level. For the rest of the paper this configuration would be denoted as a 4-2 solution.
Equivalently, the 8-term adder of Fig.~\ref{f:tree-align-add}(a) would be denoted as a 2-2-2 solution highlighting the radix of the operators used in each level of the tree.

A radix-4 operator effectively follows the baseline architecture shown in Fig.~\ref{f:serial-align-add} for 4 inputs, i.e., it finds first the maximum of the 4 exponents and subtracts it from all input exponents. The exponent differences are used  for aligning the 4 fractions before adding them. 
In fact, the proposed approach is a generalization of the baseline alignment and addition. 
The baseline approach for an $N$-term adder, shown in Fig.~\ref{f:serial-align-add}, is effectively a sub-solution of the proposed approach and uses a \emph{single} radix-$N$ operator.

\section{Evaluation}
\begin{figure}[th]
    \centering
    \includegraphics[width=0.45\textwidth]{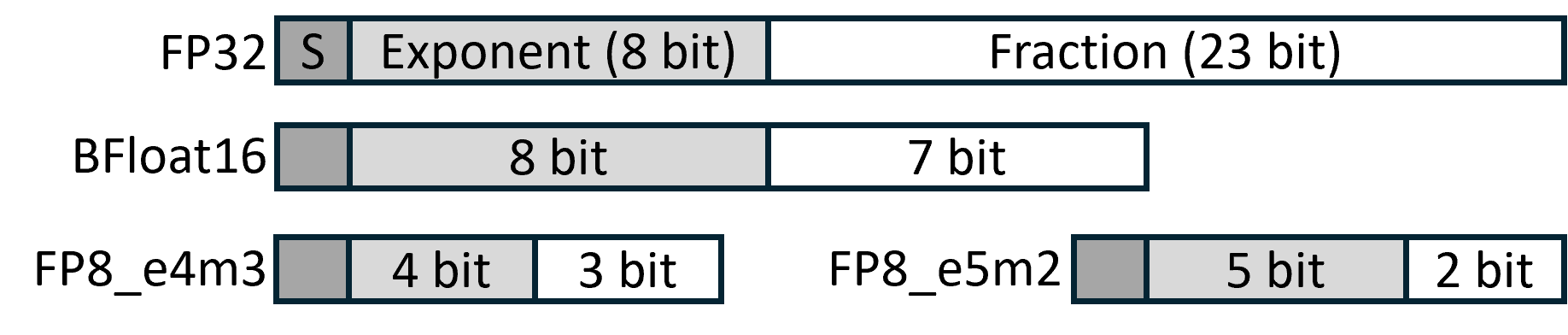}
    \caption{Structure of commonly used FP data types.}
    \label{f:fp_datatypes}
\end{figure}

Experimental evaluation aims at exploring the effectiveness of the proposed alignment and addition architecture, for building multi-term fused FP adders relative to the widely-used baseline approach. For this reason, we implemented 16, 32 and 64-term adders for the four FP-arithmetic formats shown in Fig.~\ref{f:fp_datatypes} covering single and reduced-precision formats~\cite{miniFloat}. For the proposed designs, for each multi-term adder we explored all possible configurations using align-and-add operators of various radices (i.e., number of inputs).

All the multi-term FP adders under comparison, were implemented in C++\footnote{available at {\tt github.org/ic-lab-duth/online-fp-add.git}}
and synthesized to Verilog using Catapult HLS, using a 28-nm standard-cell library. All designs (i.e., proposed and \textit{baseline}) 
operate at a clock frequency of 1 GHz and implement a complete multi-term fused FP addition that includes fraction alignment and addition as well as normalization and rounding of the final sum.
To achieve the target clock frequency, HLS synthesis was instructed to produce designs with appropriate pipeline depth, depending on the number of input terms and their data type. As the number of input terms increases so does the design's pipeline depth. For an $N$-term FP32 adder we aimed for $\log_2 N$ pipeline stages. 
HLS can derive many other pipelined solutions. However, to simplify comparisons across designs, we selected the same configuration for all cases.
For lower-precision data types, such as BFloat16 and FP8, one pipeline stage less relative to FP32 is enough to reach the targeted clock frequency due to smaller mantissa and exponent bit widths. 
The final area results were derived from Oasys logic synthesis tool. The power consumption was estimated after synthesis using the PowerPro power analysis and optimization tool. For power estimation, we employed multi-term adders in matrix multiplication kernels for the BERT Transformer~\cite{bert} using input data from the GLUE dataset~\cite{glue}. 

\begin{figure}[t]
\centering
\includegraphics[width=\columnwidth]{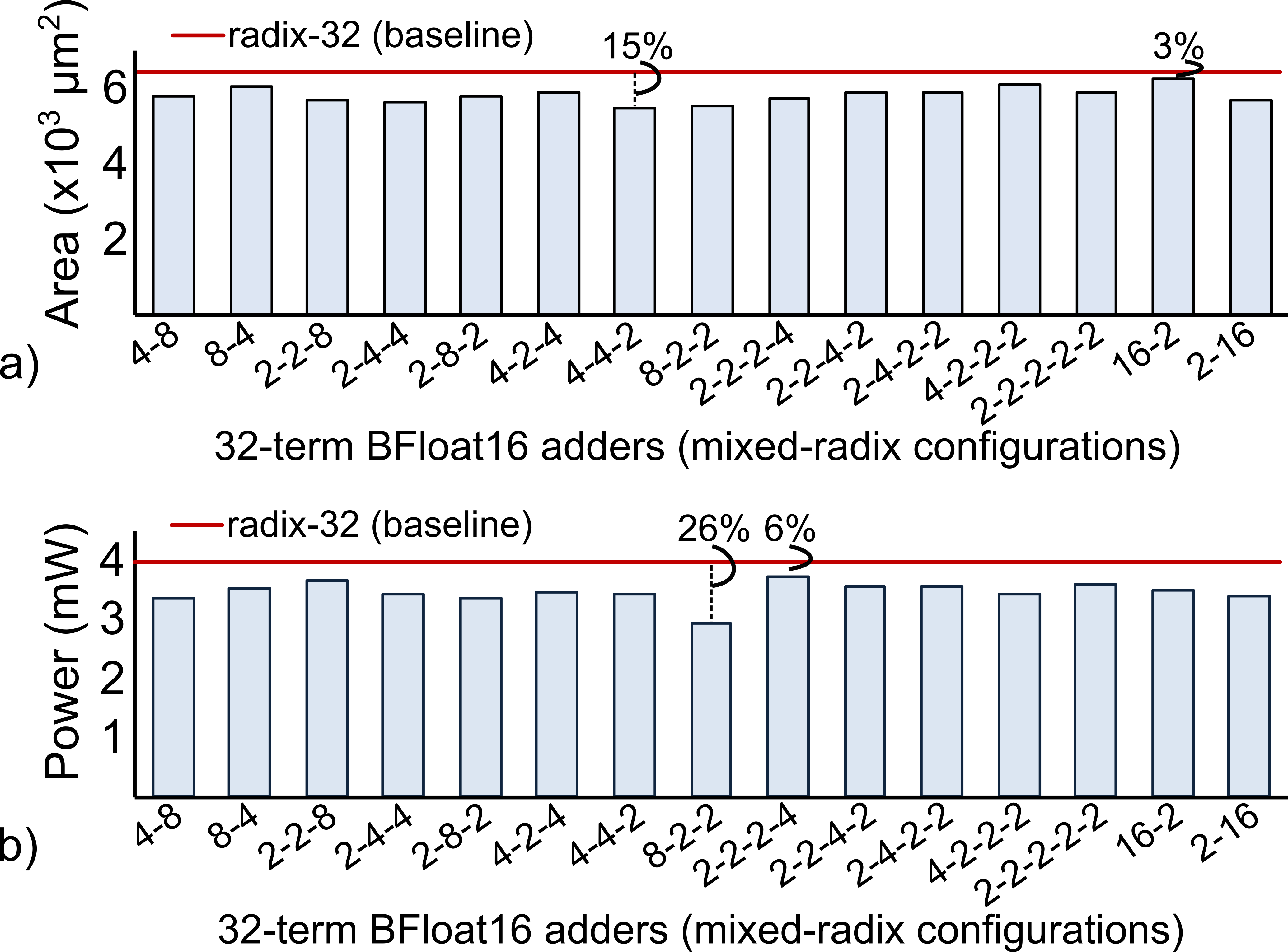}
\caption{The a) area and b) average power of 32-term BFloat16 adders designed with the baseline approach and the proposed approach that uses the newly introduced align-and-add operator $\circledcirc$ in various mixed-radix configurations.}
\label{f:arpow_1ghz}
\end{figure}

\subsection{Design-space exploration for 32-term BFloat16 adders}
\label{c:dse}
In order to assess how mixed-radix configurations perform relative to the baseline align-and-add approach in multi-term floating-point adders,
we initially focused on the case of $32$-term BFloat16 adders.
The designs presented represent complete multi-term floating point adders and the baseline approach differs from the proposed designs only in the alignment and addition logic. Normalization and rounding are the same for all designs under comparison.

Fig.~\ref{f:arpow_1ghz} depicts the area and power of the proposed adders, that follow different mixed-radix configurations relative to the baseline approach, which effectively uses a single $N$-input operator.
In all cases, utilizing a mixed-radix configuration proves more efficient than the radix-32 \textit{baseline} configuration. From the results shown in Fig.~\ref{f:arpow_1ghz}(a) the proposed designs can achieve area savings that range between 3\% and 15\%.
The 4-4-2 configuration offers the best area efficiency, reducing area by 15\%. As shown in Fig.~\ref{f:arpow_1ghz}(b), the proposed mixed-radix designs achieve power reductions of 6\% to 26\%. The optimal configuration, in terms of power consumption, is the 8-2-2 design, achieves a notable 26\% power reduction.

The proposed formulation splits alignment and addition 
to smaller hardware blocks thus increasing hardware modularity. In effect, this transformation, allows HLS to schedule intermediate alignment and addition steps to pipeline stages with better flexibility that results in more efficient designs.

\begin{figure}[t]
\centering
\includegraphics[width=0.65\linewidth]{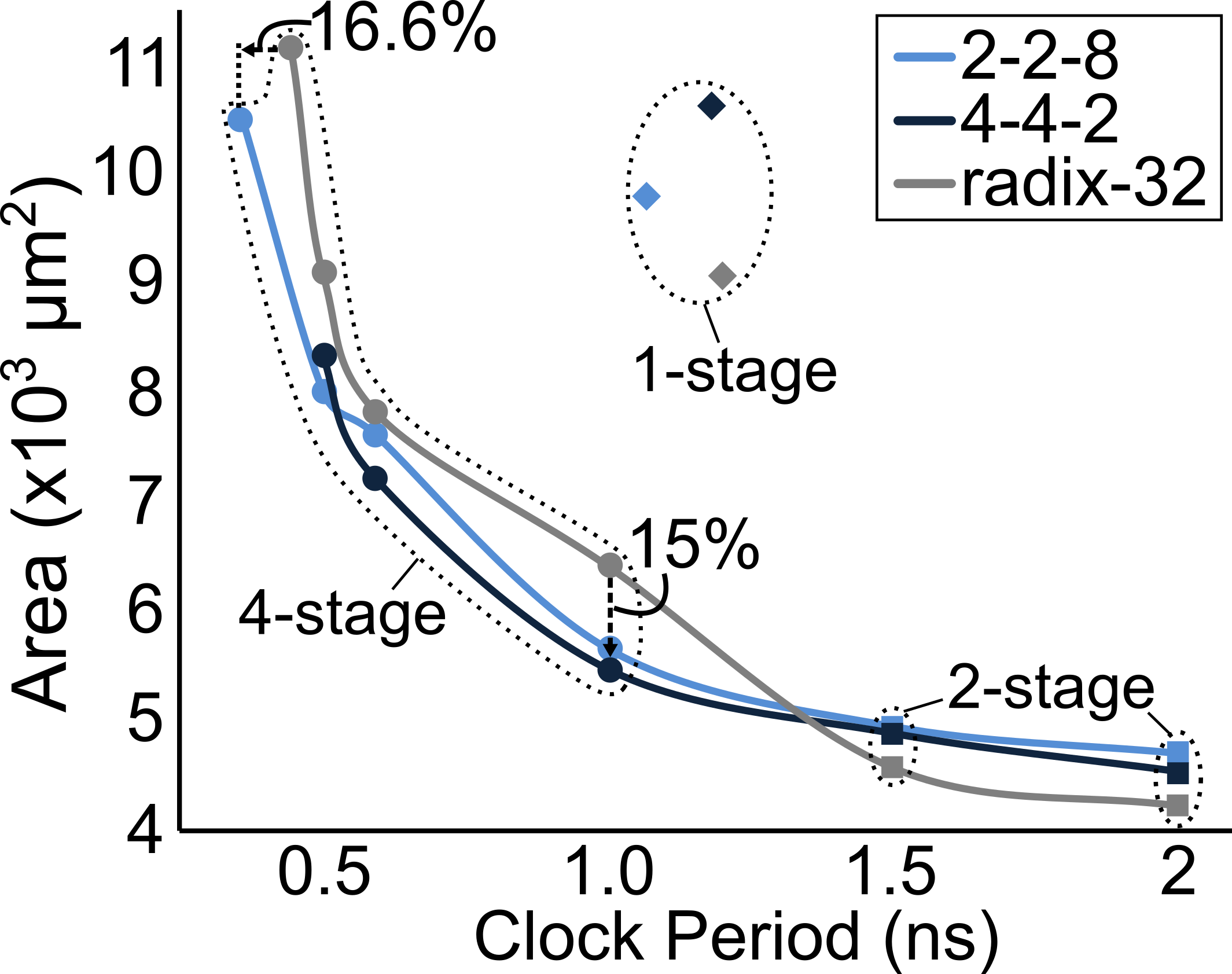}
\caption{The most area efficient designs achieved by each configuration for 32-term BFloat16 for various clock period targets using 1--4 pipeline stages.}
\label{f:lat_exp}
\end{figure}

This modular approach enhances also the delay characteristics of multi-term adders across different pipelined configurations. Fig.~\ref{f:lat_exp} illustrates the most area-efficient 32-term BFloat16 adders produced for various clock period targets. For high-frequency applications, 4-stage pipelines excel, while 2-stage pipelines are more area-optimal at lower frequencies.
The proposed 2-2-8 configuration stands out for its speed, offering a 16.6\% faster clock cycle than the baseline design with the same number of pipeline stages. In terms of area, similar to Fig.~\ref{f:arpow_1ghz}(a), the 4-4-2 design is the most compact at 1 ns. However, for less stringent clock requirements, the baseline design provides the best area-performance trade-off.
For completeness, we also included the fastest single-cycle (1-stage) implementations for each design. In all cases, their equivalent pipelined solutions offer a superior combination of speed and area efficiency for 32-term adders.

\begin{table}
\begin{center}
\caption{The area and power for (a) 16, (b) 32 and (c) 64-input multi-term adders and for various FP data types.}
\label{t:dtype_best}
\begin{adjustbox}{width=\columnwidth}
\begin{tabular}{|c||c|c|c||c|c|c|}
    \hline
        \multirow{2}{*}{\textbf{N = 16}} & \multicolumn{3}{c||}{\textbf{Area ($\times10^3\mu m^2$)}} & \multicolumn{3}{c|}{\textbf{Power (mW)}} \\\cline{2-7}
        ~ & Base & Proposed & Save & Base & Proposed & Save \\ \hline
        \multirow{2}{*}{FP32} & \multirow{2}{*}{8.87} & 6.8 & \multirow{2}{*}{23\%} & \multirow{2}{*}{3.03} & 2.65 & \multirow{2}{*}{13\%} \\
        ~ & ~ & \footnotesize(8-2) & ~ & ~ & \footnotesize(8-2) & ~ \\ \hline
        \multirow{2}{*}{BFloat16} & \multirow{2}{*}{2.92} & 2.69 & \multirow{2}{*}{8\%} & \multirow{2}{*}{1.61} & 1.35 & \multirow{2}{*}{16\%} \\ 
        ~ & ~ & \footnotesize(8-2) & ~ & ~ & \footnotesize(8-2) & ~ \\ \hline
        \multirow{2}{*}{FP8\_e4m3} & \multirow{2}{*}{1.29} & 1.23 & \multirow{2}{*}{4\%} & \multirow{2}{*}{0.83} & 0.69 & \multirow{2}{*}{17\%} \\
        ~ & ~ & \footnotesize(8-2) & ~ & ~ & \footnotesize(8-2) & ~ \\ \hline
        \multirow{2}{*}{FP8\_e5m2} & \multirow{2}{*}{1.17} & 1.23 & \multirow{2}{*}{-5\%} & \multirow{2}{*}{0.62} & 0.70 & \multirow{2}{*}{-13\%} \\
        ~ & ~ & \footnotesize(2-4-2) & ~ & ~ & \footnotesize(2-4-2) & ~ \\ \hline
        \multirow{2}{*}{FP8\_e6m1} & \multirow{2}{*}{1.33} & 1.36 & \multirow{2}{*}{-2\%} & \multirow{2}{*}{0.49} & 0.54 & \multirow{2}{*}{-10\%} \\
        ~ & ~ & \footnotesize(4-2-2) & ~ & ~ & \footnotesize(4-2-2) & ~ \\ \hline 
\end{tabular}
\end{adjustbox}
\vskip 0.2cm
\small (a) 16-term FP adders\vskip 0.2cm 

\begin{adjustbox}{width=\columnwidth}
\begin{tabular}{|c||c|c|c||c|c|c|}
\hline
    \multirow{2}{*}{\textbf{N = 32}} & \multicolumn{3}{c||}{\textbf{Area ($\times10^3\mu m^2$)}} & \multicolumn{3}{c|}{\textbf{Power (mW)}} \\\cline{2-7}
    ~ & Base & Proposed & Save & Base & Proposed & Save \\ \hline
    \multirow{2}{*}{FP32} & \multirow{2}{*}{16.24} & 14.02 & \multirow{2}{*}{14\%} & \multirow{2}{*}{6.69} & 5.78 & \multirow{2}{*}{14\%} \\
    ~ & ~ & \footnotesize(2-2-2-2-2) & ~ & ~ & \footnotesize(2-2-2-2-2) & ~ \\ \hline
    \multirow{2}{*}{BFloat16} & \multirow{2}{*}{6.44} & 5.5 & \multirow{2}{*}{15\%} & \multirow{2}{*}{3.97} & 2.92 & \multirow{2}{*}{26\%} \\ 
    ~ & ~ & \footnotesize(8-2-2) & ~ & ~ & \footnotesize(8-2-2) & ~ \\ \hline
    \multirow{2}{*}{FP8\_e4m3} & \multirow{2}{*}{3.02} & 2.51 & \multirow{2}{*}{17\%} & \multirow{2}{*}{1.85} & 1.53 & \multirow{2}{*}{17\%} \\
    ~ & ~ & \footnotesize(8-2-2) & ~ & ~ & \footnotesize(8-2-2) & ~ \\ \hline
    \multirow{2}{*}{FP8\_e5m2} & \multirow{2}{*}{2.73} & 2.44 & \multirow{2}{*}{11\%} & \multirow{2}{*}{1.74} & 1.44 & \multirow{2}{*}{17\%} \\
    ~ & ~ & \footnotesize(8-2-2) & ~ & ~ & \footnotesize(8-2-2) & ~ \\ \hline
    \multirow{2}{*}{FP8\_e6m1} & \multirow{2}{*}{2.80} & 2.48 & \multirow{2}{*}{11\%} & \multirow{2}{*}{0.76} & 0.63 & \multirow{2}{*}{18\%} \\
    ~ & ~ & \footnotesize(8-2-2) & ~ & ~ & \footnotesize(8-2-2) & ~ \\ \hline 
\end{tabular}
\end{adjustbox}
\vskip 0.2cm
(b) 32-term FP adders\vskip 0.2cm 

\begin{adjustbox}{width=\columnwidth}
\begin{tabular}{|c||c|c|c||c|c|c|}
\hline
    \multirow{2}{*}{\textbf{N = 64}} & \multicolumn{3}{c||}{\textbf{Area ($\times10^3\mu m^2$)}} & \multicolumn{3}{c|}{\textbf{Power (mW)}} \\\cline{2-7}
    ~ & Base & Proposed & Save & Base & Proposed & Save \\ \hline
    \multirow{2}{*}{FP32} & \multirow{2}{*}{32.51} & 28.67 & \multirow{2}{*}{12\%} & \multirow{2}{*}{13.26} & 10.82 & \multirow{2}{*}{19\%} \\
    ~ & ~ & {\footnotesize (2-2-2-2-4)} & ~ & ~ & {\footnotesize (2-2-2-2-4)} & ~ \\ \hline
    \multirow{2}{*}{BFloat16} & \multirow{2}{*}{12.84} & 11.73 & \multirow{2}{*}{9\%} & \multirow{2}{*}{7.30} & 7.05 & \multirow{2}{*}{4\%} \\ 
    ~ & ~ & \footnotesize(2-4-2-2-2) & ~ & ~ & \footnotesize(2-4-2-2-2) & ~ \\ \hline
    \multirow{2}{*}{FP8\_e4m3} & \multirow{2}{*}{5.79} & 5.09 & \multirow{2}{*}{12\%} & \multirow{2}{*}{3.62} & 3.01 & \multirow{2}{*}{17\%} \\
    ~ & ~ & \footnotesize (8-4-2) & ~ & ~ & \footnotesize(8-4-2) & ~ \\ \hline
    \multirow{2}{*}{FP8\_e5m2} & \multirow{2}{*}{5.34} & 4.78 & \multirow{2}{*}{11\%} & \multirow{2}{*}{3.35} & 2.78 & \multirow{2}{*}{17\%} \\
    ~ & ~ & \footnotesize(8-8) & ~ & ~ & \footnotesize(8-8) & ~ \\ \hline
    \multirow{2}{*}{FP8\_e6m1} & \multirow{2}{*}{5.39} & 4.86 & \multirow{2}{*}{10\%} & \multirow{2}{*}{1.62} &  1.35 & \multirow{2}{*}{17\%} \\
    ~ & ~ & \footnotesize(2-8-4) & ~ & ~ & \footnotesize(2-8-4) & ~ \\ \hline 
    \end{tabular}
\end{adjustbox}
\vskip 0.2cm
(c) 64-term FP adders
\end{center}
\end{table}

\subsection{Multi-term adders for various FP formats}

As previously demonstrated, the proposed approach
performs well for building 32-term BFloat16 adders.
Nevertheless, it is essential to verify that this efficiency extends to adders with fewer or more inputs and to other FP data types. A more extensive analysis will offer a comprehensive understanding of the effectiveness of FP adders built using the proposed parallel align-and-add architecture.

Table~\ref{t:dtype_best} presents the area and power performance of all designs under comparison for 16, 32 and 64 inputs and for the FP formats shown in Fig.~\ref{f:fp_datatypes}.
To examine also a corner case, where the exponent differences are large relative to the mantissa's bit width, we included also an additional 8-bit FP datatype {\tt FP8\_e6m1}.

For the proposed designs, we only report the configuration with the best area/power performance. The selected configuration is indicated inside the parenthesis below the results of the proposed designs.

The performance gains achieved by the proposed designs depend mainly on the number of input terms and are consistent across all examined FP data types. 
As shown in Table~\ref{t:dtype_best}, adders with a large number of input terms, like 32 or 64, demonstrate a more pronounced benefit compared to those with a lower number of inputs. The size of the exponent field also influences the effectiveness of mixed-radix designs. As the size of the exponent increases, exponent calculation and fraction alignment and addition become equally critical. This convergence reduces the efficiency of interleaving maximum exponent identification and
fraction alignment and addition that is leveraged by the proposed designs. Overall, the impact of the number of input terms on performance improvements is more significant than that of larger exponent fields.

\section{Conclusions}
This work reformulates the decades-old problem of serial alignment and addition appearing in multi-term FP adders in a new \emph{online} form. The proposed computation paradigm allows maximum exponent identification, exponent subtraction, alignment shift and addition to be computed incrementally and in parallel. 
Alignment and addition logic can be structured in a tree-like structure using the newly introduced align-and-add operator. Operators with varying numbers of inputs can be employed at each level of the tree. Hardware evaluation confirms that this approach substantially reduces the complexity of alignment and addition, resulting in faster multi-term FP adders or designs with smaller area and lower power consumption compared to conventional approaches.

\bibliographystyle{IEEEtran}
\bibliography{refs}

\end{document}